\documentclass[prl,superscriptaddress,twocolumn]{revtex4}
\usepackage{epsfig}

\newcommand{\simle}
{\raisebox{-0.75ex}[-1.5ex]{$\;\stackrel{<}{\sim}\;$}}

\newcommand{\dalt}
{\raisebox{-0.35ex}[-1.5ex]{$\;\stackrel{\leftrightarrow}{\partial}\!\!\!\;$}}
\def\d{{\partial}}
\def\s{{\sigma}}
\def\e{{\epsilon}}
\def\k{{ {\bf k} }}
\def\p{{ {\bf p} }}

\def\w{{\omega}}

\begin{document}

\def\runtitle{
Optical Conductivity and Hall Coefficient
in High-$T_{\rm c}$ Superconductors
}
\def\runauthor
Hiroshi {\sc Kontani}

\title{
Optical Conductivity and Hall Coefficient
in High-$T_{\rm c}$ Superconductors:\\
Significant Role of Current Vertex Corrections 
}

\author{
Hiroshi {\sc Kontani}
}

\address{
Department of Physics, Nagoya University,
Furo-cho, Nagoya 464-8602, Japan.
}

\date{\today}

\begin{abstract}
We study AC conductivities in high-$T_{\rm c}$
cuprates, which offer us significant information
to reveal the true electronic ground states. 
Based on the fluctuation-exchange 
(FLEX) approximation,
current vertex corrections (CVC's) are correctly 
taken into account to satisfy the conservation laws.
We find the significant role of the CVC's on the 
optical Hall conductivity in the presence of 
strong antiferromagnetic (AF) fluctuations.
This fact leads to the failure of 
the relaxation time approximation (RTA).
As a result, experimental highly unusual behaviors,
(i) prominent $\w$ and temperature dependences of 
$\{ {\rm Re, Im} \}R_{\rm H}(\w)$, and 
(ii) simple Drude form of $\theta_{\rm H}(\w)$ for 
wide range of $\w$,
are satisfactorily reproduced.
In conclusion,
both DC and AC transport phenomena in (slightly under-doped)
high-$T_{\rm c}$ cuprates can be explained comprehensively 
in terms of nearly AF Fermi liquid,
if one take the CVC's into account.
\end{abstract}

\sloppy

\maketitle


Non-Fermi-liquid-like behavior in 
various transport phenomena
in high-$T_{\rm c}$ superconductors (HTSC's)
is a key clue for solving the mystery of the ground state.
For example, 
$R_{\rm H}\propto T^{-1}$ is observed in hole-doped systems
below $T_0\sim 700$K, and $|R_{\rm H}|\gg 1/ne$
($n$ being the electron density) at lower temperatures
 \cite{Satoh}.
Moreover, the magnetoresistance
follows the relation
$(\Delta\rho/\rho_0)\cdot \rho_0^{2} \propto R_{\rm H}^2\propto T^{-2}$,
which is called the modified Kohler's rule
 \cite{Kimura,Ando}.
A comprehensive understanding for them cannot be achieved
by the relaxation time approximation (RTA)
even if one assume a huge anisotropy of 
the relaxation time, $\tau_\k$:
Although $R_{\rm H}\gg 1/ne$ is derived,
then $\Delta\rho/\rho_0$ increases much faster than 
$T^{-4}$ so the modified Kohler's rule is broken down
 \cite{Ioffe,Kontani-MR}.

The modified Kohler's rule is well reproduced
if one take the {\it vertex correction
for the total current ${\bf J}_\k$},
which is called the {\it back-flow}
in Landau-Fermi liquid theory
 \cite{Kontani-Hall}.
Due to the back-flow,
${\bf J}_\k$ becomes totally different
from the quasiparticle velocity ${\bf v}_\k$
in nearly AF Fermi liquid.
Reflecting this fact,
relations $R_{\rm H}\propto \xi^2\propto T^{-1}$ and
$\Delta\rho\cdot\rho_0\propto \xi^4\propto T^{-2}$ are derived,
where $\xi$ is the antiferromagnetic (AF) correlation length.
In the same way, anomalous behaviors of 
thermoelectric power and Nernst coefficient 
can be explained {\it in the unified way}
 \cite{Kontani-MR,Kontani-S,Kontani-N}.


AC transport phenomena give us further significant
information on the electronic states.
In HTSC's,
the extended-Drude (ED) relation given by the RTA,
\begin{eqnarray}
\s_{xy}^{0}(\w) &\propto& \{\s^{0}(\w)\}^2
\propto (2\gamma_{0}(\w)-iz^{-1} \w)^{-2},
 \label{eqn:ED} \\
\gamma_{0}^{-1}(\w)&=& 
 \frac2{\w} \int_{0}^{\w}d\e \langle 
 [\gamma_\k(\e-\w)+\gamma_\k(\e)]^{-1} \rangle_{\rm FS}
 \label{eqn:g0}
\end{eqnarray}
is strongly violated even when $\w \ll \gamma_{0}$
(in the far-infrared region, $\w\sim 50{\rm cm}^{-1}$)
 \cite{Drew96}.
$\gamma_\k(\w)$ is the imaginary part of the 
self-energy 
and $z(<1)$ is the mass-renormalization factor.
The ED-relation seems to hold well in Cu and Au
 \cite{DrewCuAu}.
Note that we obtain a simple Drude form
if one drop the $\w$-dependence of $\gamma_0(\w)$.
Another unexpected experimental fact is that
the optical Hall angle 
$\theta_{\rm H}(\w)\equiv \s_{xy}(\w)/\s(\w)$
in HTSC's follows a simple Drude form
even when $\w \gg \gamma_{0}$
(in the infrared region, $\w\sim 1000{\rm cm}^{-1}$),
despite the fact that $\w$-dependence of 
$\tau_\k(\w) \equiv 1/2\gamma_\k(\w)$
violated the Drude form of $\s(\w)$ 
 \cite{Drew00,Drew04}.
Previous theoretical works based on the RTA
unable to give a comprehensive understanding for them
 \cite{Drew00,Drew04}.
Thus, anomalous AC transport phneomena
have been a long-standing challenge for theory.

In this article,
we develop the method of calculating
$\s(\w)$ and $\s_{xy}(\w)$ using the FLEX approximation
by taking the current vertex correction (CVC)
to satisfy the conservation laws.
We call it the FLEX+CVC approximation.
We find that $\s_{xy}(\w)$ deviates from the ED-form
predominantly due to CVC 
when AF fluctuations are strong.
We succeeded in reproducing the
unexpected experimental behaviors of $R_{\rm H}(\w)$ 
as well as $\theta_{\rm H}(\w)$ {\it at the same time}.
Thus, anomalous AC and DC transport phenomena in HTSC's
can be understood in terms of nearly AF Fermi liquid,
if one take the CVC into account.

First, we study the self-energy $\Sigma_\k(\w_n)$ in the 
square-lattice Hubbard model using the FLEX approximation,
which is one of the self-consistent spin-fluctuation theory
 \cite{Bickers}.
This method can reproduce characteristic electronic
properties in slightly under-doped systems
above the pseudo-gap temperature
 \cite{Yamada-rev,Moriya-rev}.
We put $(t,t',t'';U)=(1,-0.1,0.1; 5)$ and $n=0.9$,
where $t$, $t'$ and $t''$ are hopping integrals
between the nearest neighbor, the second nearest and the third 
nearest, respectively.
The dispersion of conduction electron is
$\e_\k=-2t(\cos k_x + \cos k_y) -4t'\cos k_x \cos k_y
 - 2t''(\cos 2k_x + \cos 2k_y)$.
This set of parameters corresponds to
slightly under-doped LSCO.
We note that Im$\Sigma_\k(-i\delta)$ on the FS 
takes the minimum value around $(\pi/2,\pi/2)$,
which is called the 'cold-spot' in literature
 \cite{Pines-Hall}.
Its maximum (minimum) value
on the FS is 0.46 (0.081) at $T=0.02$:
The ratio of anisotropy is 5.7,
which is too small to explain the enhancement of $R_{\rm H}$;
the CVC around the cold-spot magnifies $R_{\rm H}$
 \cite{Kontani-Hall}.

Next, we study transport coefficients
in the framework of the conservation approximation,
by taking the CVC into account
 \cite{Baym-Kadanoff}.
According to the Kubo formula,
\begin{eqnarray}
\s_{\mu\nu}(\w)=\frac1{i\w}\left[
 K_{\mu\nu}^{\rm R}(\w)- K_{\mu\nu}^{\rm R}(0) \right] ,
 \label{eqn:Kubo}
\end{eqnarray}
where $K_{\mu\nu}^{\rm R}(\w)$ is the retarded
correlation function, which is
given by the analytic continuations of
the following thermal Green functions with $\w_l\ge0$
 \cite{Eliashberg,Kohno}
using the numerical Pade approximation:
\begin{eqnarray}
K_{xx}(i\w_l)
 &=& -2e^2 T\sum_{n,\k} v_{\k x}^0 g_\k^{n;l}
  \Lambda_{\k x}^{n;l} ,
 \label{eqn:sig} \\
K_{xy}(i\w_l)
 &=& i\cdot e^3 T\sum_{n,\k} \sum_{\mu,\nu}^{x,y} 
 \left[ G_\k^{n}\dalt_\mu  G_\k^{n+l} \right]
 \nonumber \\
& &\times \left[ \Lambda_{\k x}^{n;l}
 \dalt_\nu \Lambda_{\k y}^{n;l} \right]\cdot \e_{\mu\nu z},
 \label{eqn:sxy}
\end{eqnarray}
where $[A\dalt_\mu B]\equiv A\frac{\d}{\d k_\mu} B 
-  B\frac{\d}{\d k_\mu} A$, and
$\e_{\mu\nu\lambda}$ is an antisymmetric tensor 
with $\e_{xyz}=1$.
$\w_l\equiv 2\pi T l$ is a Matsubara frequency;
here we promise $l$ represents a integer and 
$n$ is a half-integer, respectively.
$G_\k^{n}\equiv G_\k(\w_n)$ is the Green function and
$g_\k^{n;l}\equiv G_\k^n G_\k^{n+l}$.
$v_{\k \mu}^0\equiv \d \e_\k / \d k_\mu$ is the
velocity of a free electron, and 
${\bf \Lambda}_{\k}^{n;l}$ is the dressed current.
In the FLEX approximation, it is given by
\begin{eqnarray}
{\bf \Lambda}_{\k}^{n;l}
 &=& {\bf v}_{\k}^0 + T\sum_{n',\p}
 V_{\k-\p}(\w_n-\w_{n'}) \cdot g_\p^{n';l}
 {\bf \Lambda}_{\p}^{n';l}
 \label{eqn:BS} 
\end{eqnarray}
where $V_{\k}(\w)$ is given in eq.(9)
of ref.\cite{Kontani-Hall}.
In eq. (\ref{eqn:BS}), we take the infinite series of the 
Maki-Tompson terms in ${\bf \Lambda}$, that is,
we drop all the Aslamazov-Larkin terms.
This simplification is justified for DC-conductivities
when the AF fluctuations for ${\bf Q}\approx(\pi,\pi)$
are dominant, as proved in ref.\cite{Kontani-Hall}.
This will be also true for optical conductivities,
due to the fact that 
the $f$-sum rules both for $\s(\w)$ and $\s_{xy}(\w)$
are well satisfied in the present study,
as will be discussed below.

\begin{figure}
\begin{center}
\epsfig{file=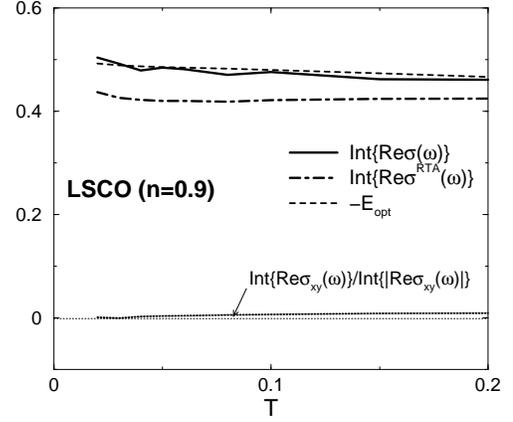,width=6.5cm}
\end{center}
\caption{
$f$-sum rules both for $\s(\w)$ and $\s_{xy}(\w)$
are well satisfied if the CVC is correctly taken 
into account.
This fact assures the reliability of the
present numerical study.
}
  \label{fig:sumrule}
\end{figure}

Pade approximation is less reliable
when the function under consideration
is strongly $\w$-dependent,
and when the temperature is high because
the Matsubara frequency is sparse.
To obtain reliable results,
both $\Sigma_\k(\w_n)$ and ${\bf \Lambda}_\k^{n;l}$ 
have to be converged so that the relative errors
should be under $10^{-8}$.
Moreover, we utilize the fact that the $i\w$-linear term
of $K_{\mu\nu}(\w)$ is equal to the DC value of $\s_{\mu\nu}$,
which can be obtained by the FLEX+CVC approximation
with high accuracy, as performed in refs.
\cite{Kontani-Hall,Kontani-MR,Kontani-S,Kontani-N}.
As a result, we succeed in deriving the
$\s_{\mu\nu}(\w)$ for any $\w$ with enough accuracy.

To confirm the accuracy of numerical results,
we check the following $f$-sum rules
 \cite{Kubo}:
\begin{eqnarray}
\int_0^\infty d\w {\rm Re}\s(\w)
&=& \pi e^2 \sum_{k} \frac {\d^2 \e_\k}{\d k_x^2}n_\k ,
 \label{eqn:sum-sig}
 \\
\int_0^\infty d\w {\rm Re}\s_{xy}(\w)
&=& 0 .
 \label{eqn:sum-sxy}
\end{eqnarray}
Because the FLEX+CVC approximation 
is a conserving approximation,
$f$-sum rules should be satisfied in the present study
if numerical error is absent.
In Fig.\ref{fig:sumrule},
Int$\{\s(\w)\}$ (Int$\{\s_{xy}(\w)\}$) represents
the numerical result for the left-hand-side of 
eq. (\ref{eqn:sum-sig}) (eq. (\ref{eqn:sum-sxy})),
and $-E_{\rm opt}$ is that for the right-hand-side
of eq. (\ref{eqn:sum-sig}).
We see that the $f$-sum rules hold well, 
whose relative error is below $2.5\%$.
This results assure the high reliability
of the present numerical study,
considering the ill accuracy of the Pade approximation
for larger $|\w|$.
On the other hand,
Int$\{{\rm Re}\s^{\rm RTA}(\w)\}$ (without any CVC) 
is at least 12$\%$ smaller than the correct value,  
reflecting the violation
of the conservation laws in the RTA.

\begin{figure}
\begin{center}
\epsfig{file=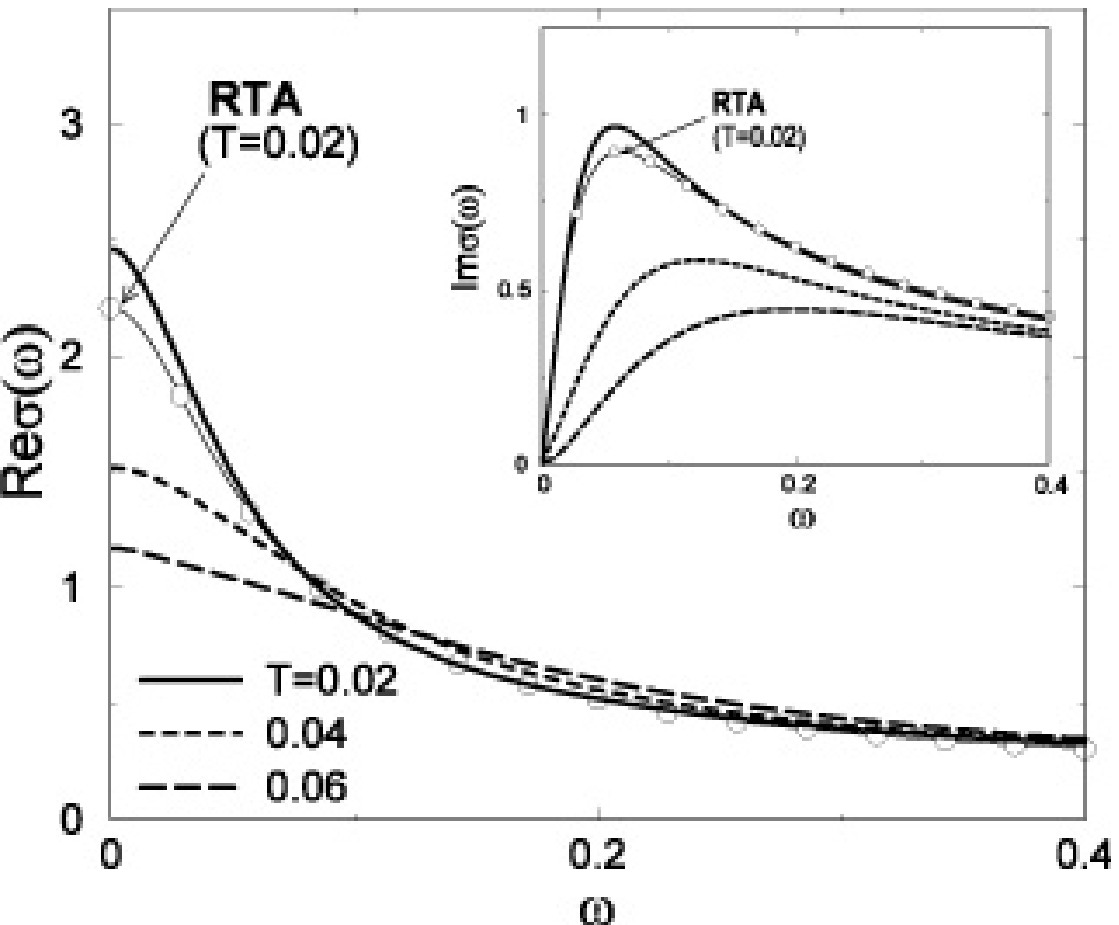,width=7cm}
\epsfig{file=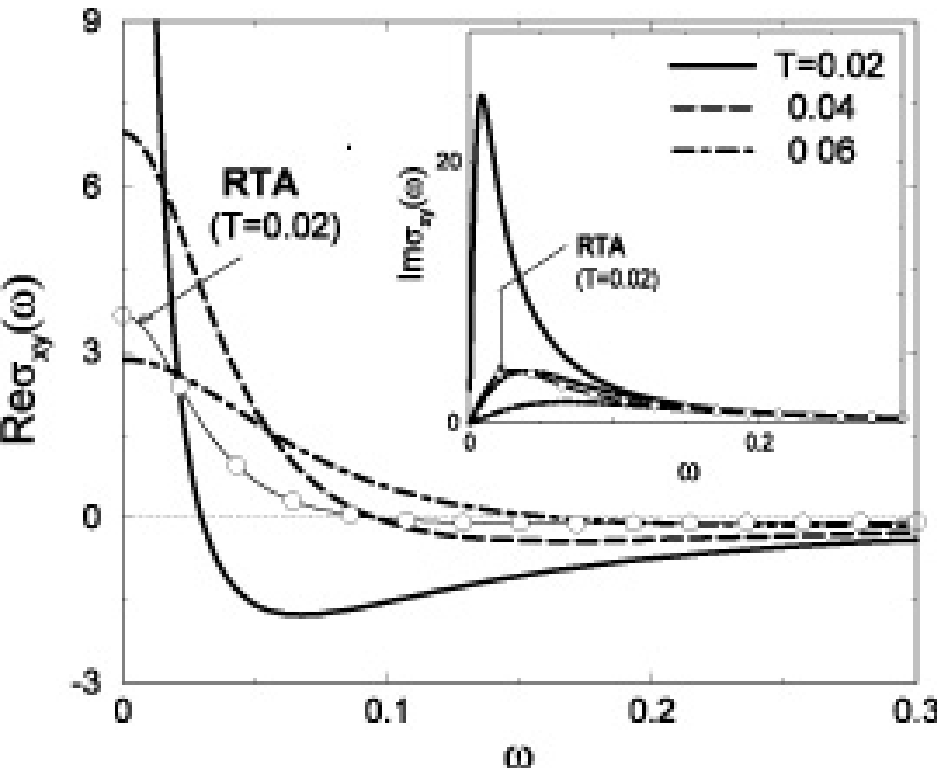,width=7cm}
\epsfig{file=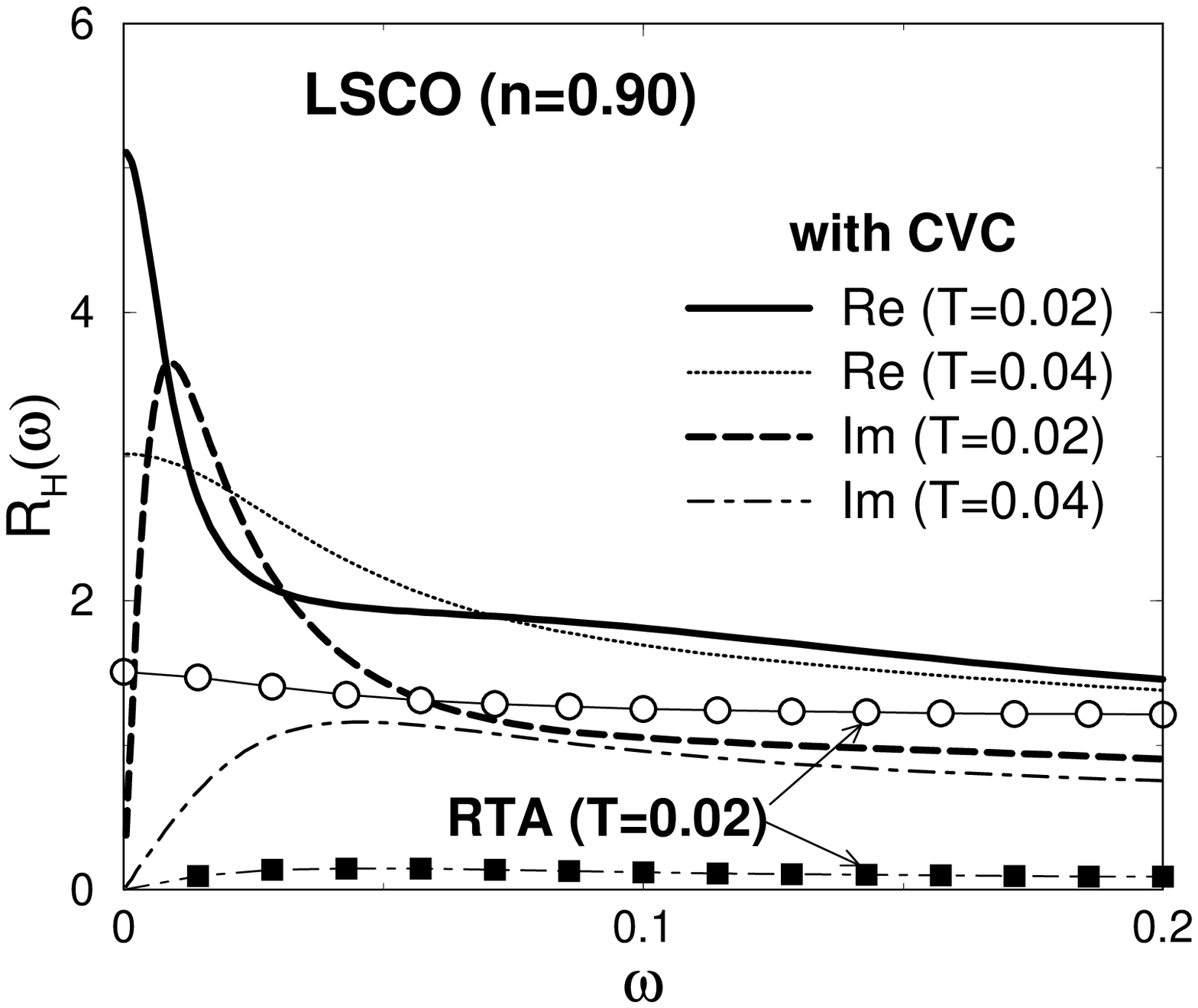,width=7cm}
\end{center}
\caption{
$\s(\w)$, $\s_{xy}(\w)$ and $R_{\rm H}(\w)$
by the FLEX+CVC approximation (with CVC).
$T=0.02$ and $\w=0.1$ approximately 
correspond to 80K and 300${\rm cm}^{-1}$,
respectively. 
}
  \label{fig:opt}
\end{figure}

Figure \ref{fig:opt} shows
the $\w$-dependences of $\s(\w)$, 
$\s_{xy}(\w)$ and $R_{\rm H}(\w)$ for several temperatures.
For comparison, we also plot the results by the RTA 
($\s^0(\w)$, $\s_{xy}^0(\w)$ and $R_{\rm H}^0(\w)$)
at $T=0.02$, which is given by dropping all the CVC's.
We have checked that both $\s^0(\w)$ and $\s_{xy}^0(\w)$
satisfy the ED form in eq.(\ref{eqn:ED})
for $\w\simle0.2$
 \cite{future}.
As shown in Fig.\ref{fig:opt},
the CVC for $\s(\w)$ is less effective
(at least within the FLEX approximation),
as is the case of the DC $\s$ 
 \cite{Kontani-Hall}.
This fact suggests that $\s(\w)$ follows the ED-form well.
One can see that Re$\s(\w)$ decreases with $\w$ 
much slower than the Lorentzian, because 
$\gamma_0(\w)$ is an increase function of $\w$.

In the Fermi liquid theory, 
the CVC is divided into (i) the back-flow, and 
(ii) the renormalization of ${\bf v}_\k$
by $d\Sigma_\k/dk$.
In nearly AF systems,
the conductivity is reduced by the former,
whereas it is enhanced by the latter
 \cite{Kontani-Hall}.
As shown in  Fig.\ref{fig:opt},
The latter slightly exceeds the former
in the present calculation.

In contrast,
$\s_{xy}(\w)$ with full CVC is quite different from 
the RTA's result:  Both $\w$ and temperature dependences
are much enhanced by CVC.
Re$\s_{xy}(\w)$ at $T=0.02$ 
takes a large negative value for 
$\w>0.03\sim 120{\rm cm}^{-1}$,
which is consistent with experimental observations
 \cite{Drew04,Drew00,Drew96}.
Although Re$\s_{xy}^0(\w)$ by RTA becomes negative
for $\w>0.1$, its absolute value is very small.
It is naturally understood from the ED-form
because $\gamma(\w)$ increases with $\w$.
This large dip in Re$\s_{xy}(\w)$ is naturally
understood in terms of the $f$-sum rule, 
eq.(\ref{eqn:sum-sxy}), because Re$\s_{xy}(0)>0$ 
takes an enhanced value due to the CVC.
We also stress that 
${\rm Im}\s_{xy}(\w)/\w|_{\w=0}$ 
is strongly enhanced due to the CVC;
Im$\s_{xy}(\w)$ at $T=0.02$ takes the maximum value
at $\w_2\sim 0.01$, which is about six times larger
than $\w_1$ for Im$\s(\w)$.

The violation of the ED-form is 
apparent in $R_{\rm H}(\w)=\s_{xy}(\w)/\s^2(\w)$.
We see that 
$R_{\rm H}^0(\w)$ by the RTA, 
$\s_{xy}^0(\w)/(\s^0(\w))^2$,
is almost $\w$-independent and its imaginary part is tiny,
which means that
both $\s_{xy}^0(\w)$ and $\s^0(\w)$ follow
the ED-forms given in eq.(\ref{eqn:ED}).
On the other hand,
$R_{\rm H}(\w)$ given by the FLEX+CVC approximation
shows prominent frequency as well as 
temperature dependences.
The obtained overall behavior
is similar to the experimental observation
in a optimally-doped YBCO at 95K
  \cite{Drew96} .
Im$R_{\rm H}(\w)$ at $T=0.02$ takes the maximum value
at $\w_3\sim 0.01$, which is similar to $\w_2$ for 
Im$\s_{xy}(\w)$ and is six times larger than 
$\w_1$ for Im$\s(\w)$.
The relation $\w_1\gg\w_2\sim\w_3$
in the present study,
which is consistent with
experimental observation \cite{Drew04,Drew96},
cannot be reproduced by the RTA:
It can be explained only when
the back-flow is taken into account
 \cite{future}.

\begin{figure}
\begin{center}
\epsfig{file=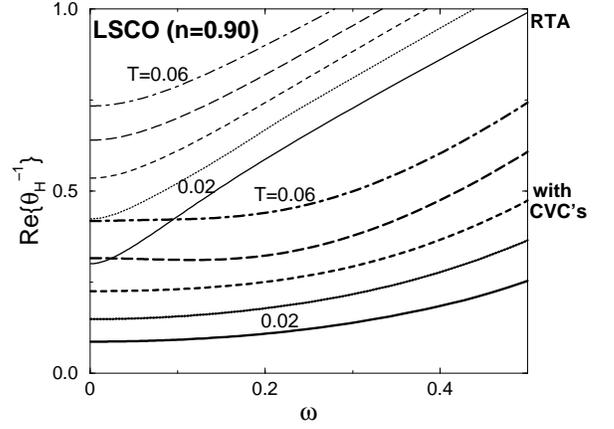,width=7.5cm}
\end{center}
\caption{
$\w$-dependences of  $\theta_{\rm H}(\w)$
given by the RTA and by the
FLEX+CVC approximation.
}
  \label{fig:IHA-w}
\end{figure}
\begin{figure}
\begin{center}
\epsfig{file=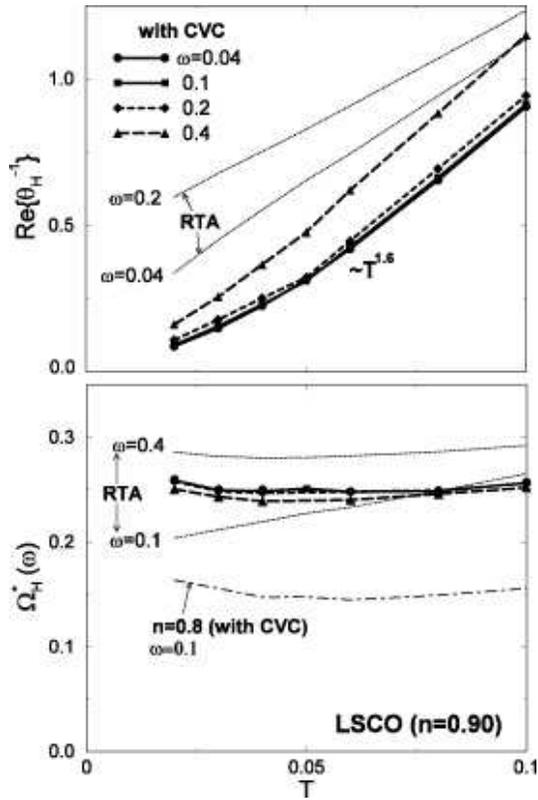,width=7cm}
\end{center}
\caption{
Temperature dependence of $\theta_{\rm H}(\w)$
by the FLEX+CVC approximation.
}
  \label{fig:IHA}
\end{figure}

IR optical Hall angle $\theta_{\rm H}(\w)=\s_{xy}(\w)/\s(\w)$
($\w\simle 1000{\rm cm}^{-1}=1440$K)
has been intensively measured by Drew et al
 \cite{Drew04,Drew00}.
They found that the following simple Drude form is 
well satisfied for the Hall angle:
\begin{eqnarray}
\theta_{\rm H}(\w)
 &=& \frac{\Omega_{\rm H}^\ast}{2\gamma_{H}^\ast-i\w} ,
 \label{eqn:D-HA}\\
\gamma_{H}^\ast &\propto& T^{-d}; \ \ d=1.5\sim2
 \nonumber \\
\Omega_{\rm H}^\ast &\propto& T^0 ,
 \nonumber 
\end{eqnarray}
where
$\gamma_{H}^\ast$ and $\Omega_{\rm H}^\ast$
are $\w$-independent for IR range ($\w\simle1000{\rm cm}^{-1}$).
This is highly nontrivial because 
$\gamma_{0}(\w)$ in eq.(\ref{eqn:g0})
is approximately linear-in-$\w$ for $\w\gg T$ in HTSC's
 \cite{Pines-opt}.
$\Omega_{\rm H}^\ast$ is almost independent
of $\w$ and $T$, which increases as the doping decreases.
These unexpected results
put very severe constraints on theories of HTSC.

From now on,
we show that the Drude-type form of the Hall angle
is reproduced quite well by the FLEX+CVC approximation.
Figure \ref{fig:IHA-w} shows the $\w$-dependence of
Re$\{ \theta_{\rm H}^{-1}(\w) \}$.
Figure \ref{fig:IHA} also shows the $T$-dependence of
Re$\{ \theta_{\rm H}^{-1}(\w) \}$ and
$\Omega_{\rm H}^\ast(\w)\equiv 
-\w/{\rm Im}\{ \theta_{\rm H}^{-1}(\w)\}$
for several $\w$'s, respectively.
They correspond to
$2\gamma_{\rm H}^\ast/\Omega_{\rm H}^\ast$ and
$\Omega_{\rm H}^\ast$, respectively,
if we apply the Drude form in eq. (\ref{eqn:D-HA}).
We can recognize that $\Omega_{\rm H}^\ast(\w)$
is almost independent of $\w$ and $T$ for
$\w \le0.4$ and $T \le 0.1$,
while it shows sizable $(\w,T)$-dependences
within the RTA.
Thus, Re$\{ \theta_{\rm H}^{-1}(\w) \}$
shown in  Fig.\ref{fig:IHA-w} and in the upper panel of Fig.\ref{fig:IHA}
represents the $(\w,T)$-dependences of $\gamma_{\rm H}^\ast$.
For $\w\le0.2$, $\gamma_{\rm H}^\ast$ by the FLEX+CVC
approximation is almost equal to its DC value
and $d=1.6$ in eq. (\ref{eqn:D-HA}).
$d$ becomes smaller for $\w=0.4$.
On the other hand,
$\gamma_{\rm H}^\ast$ by RTA is almost linear-in-$T$
and its $\w$-dependence is sizable,
which is inconsistent with experiments.
To summarize, experimental simple Drude-form of $\theta_{\rm H}(\w)$,
which cannot be understood by the RTA,
is well reproduced if one take the CVC into account.
The experimental value of 
$\Omega_{\rm H}^\ast/B$ for slightly under-doped YBCO
and BSCCO ($n\sim0.9$) is about $0.3[{\rm cm^{-1}T^{-1}}]$,
which corresponds to $0.25$ in the present energy scale.
It decreases to be
$0.15[{\rm cm^{-1}T^{-1}}]$ for optimally-doped YBCO.
Thus, the doping dependence of $\Omega_{\rm H}^\ast$,
as well as its value, are reproduced qualitatively.
Similar results are obtained 
when the set of parameters for YBCO given in 
ref.\cite{Kontani-Hall} is used.

Now, we introduce several
results given by an approximate solution
of the Bethe-Salpeter equation for ${\bf J}_{\rm k}$
at finite frequencies
 \cite{future}:
The general expressions for $\s(\w)/\w|_{\w=0}$
and $\s_{xy}(\w)/\w|_{\w=0}$ can be derived 
from the Kubo formula 
 \cite{future}.
By analysing the CVC included in them,
we obtain the relation
$\s_{xy}(\w) \approx a(2\gamma_0)^{-1}
+ b\cdot iz^{-1}\w (2\gamma_0)^{-2} + O(\w^2)$,
where $a\propto \xi^2$ and $b\propto \xi^3$
due to the CVC, whereas $a=b=1$ in the RTA.
As a result, $\s_{xy}(\w)$ deviates form the 
ED-form due to the CVC.
On the other hand,
$\s(\w) \approx (2\gamma_0)^{-1}
+ iz^{-1}\w (2\gamma_0)^{-2}$
even if the CVC is taken.
Then, we obtain relations
Im$R_{\rm H}(\w)/\w
 \propto z^{-1}b\gamma_0^{-1} \propto T^{-2.6}$ and 
Im$\theta_{\rm H}^{-1}(\w)/\w 
 \propto z^{-1}ba^{-2} \propto T^{0}$,
because 
$\xi\propto T^{-0.4}$ and $z^{-1}\propto T^{-0.4}$
in the present FLEX approximation
 \cite{future}.
These relations are confirmed by the present
numerical study.
In addition, the cancellation of the $\w$-dependences
of $\gamma_0(\w)$ and CVC's result in the almost
constant $\gamma_{\rm H}$ in eq.(\ref{eqn:D-HA}).

In summary,
we have calculated the optical conductivities for HTSC's 
by the FLEX+CVC approximation, which had been serious problems
associated intimately with the true electronic ground states 
in HTSC's.
Experimentally observed
anomalous behaviors for $\s(\w)$, $\s_{xy}(\w)$,
$\theta_{\rm H}(\w)$ and $R_{\rm H}(\w$)
are well reproduced {\it for enough wide range of 
frequencies and temperatures},
without assuming any fitting parameters.
They are consistent with the characteristic
experimental results for HTSC's reported by Drew et al.
 \cite{Drew96,Drew00,Drew04},
which cannot be reproduced by previous 
theoretical works based on the RTA
even if one assume extremely anisotropic $\tau_\k$.
The present study ensures that 
anomalous AC and DC transport phenomena in HTSC's
can be understood in terms of nearly AF
Fermi liquid state.

There remain many important issues 
for future study.
We will calculate the optical Hall conductivity 
for electron-doped HTSC's
 \cite{future}.
We are also planning to study
the optical conductivities in the pseudo-gap region
based on the FLEX+T-matrix approximation,
which ascribes the pseudo-gap phenomena in HTSC's 
to the strong superconducting fluctuations
 \cite{Kontani-N,Yamada-rev}.
We expect that the
non-Drude form of the far-IR ($\w=20\sim250{\rm cm}^{-1}$)
Hall angle in YBa$_2$Cu$_3$O$_7$ below 150K
 \cite{Drew02}
might be explained by this study.


\end{document}